# Exploring the feasibility of electric vehicle travel for remote communities in Australia

Keigan Demaria[a] and Björn C. P. Sturmberg[a*] and Brad Riley[b] and Francis Markham[b]

[a] Battery Storage and Grid Integration Program; Research School of Engineering; College of Engineering and Computer Science; The Australian National University, Canberra, Australia;

[b] Centre for Aboriginal Economic Policy Research; College of Arts and Social Sciences; The Australian National University, Canberra, Australia;

*corresponding author: bjorn.sturmberg@anu.edu.au

Keigan Demaria earned his Honours in the School of Engineering at the Australian National University.

Björn C. P. Sturmberg is a Research Leader in the Battery Storage and Grid Integration Program at the Australian National University where he conducts transdisciplinary work to accelerate the transition to decarbonised energy and transport systems.

Brad Riley is a Research Fellow at the Centre for Aboriginal Economic Policy Research at the Australian National University working on First Nations benefit in the energy transition as part of the Zero Carbon Energy for the Asia Pacific Grand Challenge.

Francis Markham is a Research Fellow at the Centre for Aboriginal Economic Policy Research at the Australian National University. His research aims to integrate critical geographic theory with quantitative methods, in particular the social applications of Geographic Information Systems.

# Exploring the feasibility of electric vehicle travel for remote communities in Australia

Remote communities in Australia face unique mobility challenges that stand to be further complicated by the transition from Internal Combustion Engine (ICE) vehicles to Electric Vehicles (EVs). EVs offer a range of advantages that include lower maintenance requirements and independence from costly, dangerous and polluting petroleum imports that have long been burdensome for remote communities. Yet the adoption of electric vehicles in Australia has been slow by international standards, and what policy strategies do exist tend to focus on incentivizing uptake among urban residents with the means to afford new technologies, potentially leaving remote communities in the 'too hard basket'. In this study we assess the feasibility of EVs for a sample of communities in remote Australia using Geographic Information System analysis of travel distances between communities and service hub towns utilizing present-day EV specifications and charging technologies. We show that while EV travel is often not currently feasible for trips to large service hub towns using low-range vehicles, over 99% of communities and residents considered would be able to travel to their nearest small service hub town with existing long-range EVs. This finding suggests that while the barriers to the electrification of transport in remote communities are significant, they are not insurmountable and are deserving of consideration in national and state policy developments in the deployment of charging infrastructure.

Keywords: electric vehicles; range anxiety; electric vehicle charging; hub and spoke model; remote communities; geographic information system; transport policy

**Introduction**

Responsible for 25% of global emissions and more than 18% of Australia's greenhouse gas pollution, the transport sector must rapidly decarbonize if we are to limit the worst effects of climate change (Climate Council 2018; Cozzi et al. n.d.; Australian Government 2020). Fortunately, the transition from Internal Combustion Engine (ICE) vehicles to Electric Vehicles (EVs) brings with it significant public and private benefits including reduced greenhouse gas emissions, reduced noise pollution and reduced health risks through improved air quality and reduced reliance on costly imported petroleum products (Broadbent, Metternicht G., and Drozdzewski 2019; Brugge, Durant, and Rioux 2007; National Transport Commission 2020). Globally the transition to electric vehicles is gathering momentum with strong government support in major markets and EVs forecast to reach price parity in total cost of ownership with ICE vehicles between 2022-2024 (IEA 2020). EV sales are expected to grow from 3% in 2020 to 7% (5.4 million vehicles) in 2023 and 58% by 2040 (Bloomberg NEF 2021; Irles 2019; IEA 2020). Yet Australia lags well behind our international peers in EV adoption (Dwyer et al. 2021; Rafique and Town 2019). In 2021 EVs accounted for only 2% of new passenger vehicles sales in Australia, compared with 15% in the UK and 72% in Norway (Paoli and Gul 2022). Australia's torpid rate of adoption can be attributed to several factors related to an absence of supportive federal government policy including the limited number of available EV models, limited public charging infrastructure as well as consumer anxiety about limited trip range (Rafique and Town 2019; Gong, Ardeshiri, and Hossein Rashidi 2020). Despite this sluggish start, government forecasts expect EVs to make up at least 70% of new vehicle sales and approximately 30% of the vehicle fleet by 2040 (Infrastructure Australia 2019a). Independent research has found that science-based targets to keep global warming to less than 1.5 degrees require even faster rates of electrification due to the long-lived nature of vehicles and the very limited remaining carbon budget (Whitehead et al. 2022). These targets include requiring a clear strategy for decarbonising all transport modes, with interim targets specified by 2025; very strong private sector transition to low emissions vehicles (cars, trucks and buses) by 2035, net-zero land transport emissions achieved by 2045 and the decarbonisation of domestic and international aviation during the 2040's as required to achieve net-zero transport sector emissions by 2050 (Whitehead et al. 2022). As evidenced by the impact of direct current fast charging deployment in Norway, investment in public charging infrastructure is correlated with high levels of EV uptake globally, while the reverse is also true in that a lack of public charging infrastructure typically limits the uptake of EVs (EV Council 2020; Broadbent, Metternicht G., and Drozdzewski 2019; Dwyer et al. 2021). In the absence of coordinated federal strategy, policies and planning, Australia's EV charging network has been developed primarily by State and Territory governments through the installation of a limited number of fast charging facilities along the more popularly travelled highways and an increasing number of privately operated charging networks (Queensland Government 2018a; Distributed Energy Integration Program 2019; Braid 2020).

The transition to electric vehicles likely holds unique opportunities and challenges for Australia's many remote communities. Relatively small and widely dispersed, remote settlements have long relied heavily upon costly road transportation for fuel, goods and services, as well as to access distant markets in larger 'hub' towns (Transport and Infrastructure Council 2015; A. Spandonide 2015a). In this context EVs may present an attractive solution for eliminating expensive imported liquid fuels while reducing maintenance and freight costs, due to fuel substitution and EVs having up to one hundred times fewer moving parts than ICE vehicles (Hunt et al. 2019; Whitby 2019; Electric Vehicle Council 2020).  In future EVs could foreseeably play an important role as an additional source of energy storage in support of the operation and resilience of renewable energy microgrids – reducing costly grid development, extension or upgrade and supplementing community batteries (Broadbent, Metternicht G., and Drozdzewski 2019; Malmgren 2016; Nelder, Newcomb, and Fitzgerald 2016). Yet the transition to electric vehicles in remote communities faces significant challenges, including; actual (and psychological) range levels of EVs (Franke and Krems 2013; Pearre et al. 2011; Steg 2005; Turrentine and Kurani 2007), unsealed and corrugated roads and a lack of all-weather access (A. Spandonide 2015a; Guerin and Guerin 2018); the availability and suitability of chargers (Dwyer et al. 2021; Infrastructure Australia 2019b; Tesla Owners Australia 2020), temperature extremes, constraints on the capacity of existing electricity supplies limiting the number and rate of charging (Tan, Ramachandaramurthy, and Yong 2016; Clement-Nyns, Haesen, and Driesen 2011; Shafiee, Fotuhi-Firuzabad, and Rastegar 2013) as well as generally lower levels of public and private investment coupled with a lack of resources to independently develop infrastructure (Transport and Infrastructure Council 2015; Sierzchula et al. 2014; Dockery 2015). These challenges should not preclude the need for detailed examination of how remote communities might soon find benefit and opportunity in the transition to electromobility.

In this paper we provide a preliminary examination of electric vehicle travel for remote communities north of the 28[th] parallel by investigating the feasibility of travel within an experimental hub-and-spoke model, whereby residents of small communities travel to larger hub towns to access essential goods and services. We acknowledge while EV driving ranges are increasing iteratively, a multiplicity of factors are likely to impact upon real-world EV range, including; the role of driving behaviour, traffic, meteorological and environmental factors (such as temperature extremes and the attendant use of air-conditioning) and road surface (NT Department of Infrastructure 2019; Oreizi n.d.). For the purpose of this experimental study, we choose to focus principally on the question of whether present-day EV trip ranges (as stated by the manufacturer) are sufficient to cover requisite distances in two assumed hub and spoke accessibility scenarios using (presently available) sample high and low range electric vehicles. We leave for future work the challenge of optimizing locations for charging infrastructure, consideration of those factors impacting upon vehicle trip ranges as well as any estimate of the capacity of existing remote power systems to cover the additional energy demands likely created by EV charging. Calculating potential demand and the implications for the cost for future energy systems (ie. solar farms/batteries onsite at remote community locations) to cover both stationary energy and transport demands is an important are for future work. These more granular details will need careful study on a case-by-case basis as conditions on community vary across jurisdictions.

**Background**

***Remote Communities***

Remote Australia accounts for 85% of the national landmass yet is sparsely populated with only 2.3% of the population (Markham and Doran 2015). For many small and widely dispersed communities in remote Australia, transport services are typically less safe and reliable and more costly than those experienced by residents of non-remote Australia, a disparity likely exacerbated in the context of changes in socio-economy and climate (A. Spandonide 2015a). Distant from primary resources and support services many remote communities experience challenges accessing essential services such as water, energy, suitable housing and adequate health and education as well as access to cellular networks and the terrestrial National Broadband Network (NBN) (Guerin and Guerin 2018; Taylor 2003; Markham and Doran 2015; Daphne Habibis and Daphne Habibis 2019). Some community access roads remain unsealed (for example approximately 75% of community roads in the Northern Territory are unsealed) which creates difficulties when inclement weather restricts access for weeks or months and impacts the ability of community members to travel to regional towns for services, to visit family and friends or to commute for work in larger centres (A. Spandonide 2015b). Building and maintaining transport infrastructure to service freight and logistics involved in the provision of food and fuels in remote communities typically leads to higher costs of goods and services (Bräunl et al. 2020). Moreover, electricity supplies in remote communities are often characterised by weak 'edge-of-grid' connections to the electricity network or 'off-grid' diesel or gas fuelled generation, which incongruously results in some of the highest marginal energy generation costs in Australia within regions otherwise host to world class renewable energy resources (Bräunl et al. 2020; Moss, Corman, and Blashki 2014; Frearson, L., Rodden, P., Backwell, J., Thwaites 2015; Pittock 2011).

Remote and regional communities currently manage requisite refuelling for their journeys from a combination of hub and highway petrol stations and en-route roadhouses who typically exploit a variety of strategic significance in their location (Spandonide 2017). Examples include Pardoo which provides a welcome en-route stop between Port Hedland and Broome and is located at the intersection of the Great Northern Highway and the Shay Gap Road. Timouth Well Roadhouse at the base of the Tanami Road (862km south of Halls Creek) provides refuelling both for those transiting from the Kimberley to Central Australia and resident remote communities in both the Northern Territory and Western Australia (Fuel Map 2022). A rationalisation in the number of service stations has seen their number decline from around 20,000 in the 1970's to around 6400 (Knight Frank 2017). As elsewhere liquid fuel prices in these locations are determined by both overseas and local market forces and movements in the AUD-USD exchange rate, combined with generally higher costs for transport and fuel storage. Lower population densities and demand (including for convenience sales such as drinks, food and newspapers that can enable retailers to add to overall profits and keep fuel prices lower) result in fewer outlets and less competition (Australian Competition and Consumer Commission 2022).

***Hub and Spoke Service Delivery***

'Hubs' are typically central nodes in a network deliberately located to facilitate connectiveness, to attract a high volume of traffic from proximate 'spoke' locations so as to efficiently host, arrange, or deliver services across a range of design applications (O'Kelly 1998; Rodrigue 2020). Examples of 'hubs' include the siting of medical or aviation services so as to facilitate high volumes of patients or arriving and departing flights (House 1998; Devarakonda 2016; RASS 2017). The robustness of service hubs can enhance the efficiency of their connected network by improving performance according to a set of system specific criteria, such as; range, speed or volume of access (Elrod and Fortenberry 2017; O'Kelly 1998; Rodrigue 2020). For example, the Northern Territory government employs a hub-and-spoke system of service delivery for remote communities by using a rule-driven location-allocation to define service hubs, an arrangement which has been shown to improve network efficiency in terms of reduced road travel time (Markham and Doran 2015). Given the developmental nature of electric vehicle travel for remote communities, hub-and-spoke methodology provides one possible approach to assessing 'feasibility', assuming adequate charging infrastructure were to be developed in defined major and minor service 'hubs'- towns and smaller dependent communities between which goods, services and people flow via (potentially electrified) transportation services.

### Travel Behaviours

For the uptake of EVs in remote communities to be feasible they must satisfy current mobility practices, while planning for a transition to electric vehicles requires understanding both extant and future travel patterns to be serviced, including origin, destination, distance, frequency as well as more granular details such as stay length and motivation (Guerin and Guerin 2018). Travel behaviours can be influenced by spatial characteristics such as road networks (Tang 2019), meteorological factors (Budnitz, Chapman, and Tranos 2019) attitude, age and income (von Behren et al. 2020) and may change over time (Liu and Xu 2018). In one example in the remote Northern Territory Taylor (Taylor 2003) utilised primary data to define service centre locations and population catchments based on survey responses of "the nearest town that members of the community usually go for banking and major shopping services" to identify 35 service centres ranging from larger towns such as Darwin & Alice Springs to smaller towns like Timber Creek and Borroloola. While outdated the survey usefully indicates that ~ 15,000 people across 259 communities in the Northern Territory, South Australia & Western Australia accessed Alice Springs as their primary service centre. Similarly, the CRC for Remote Economic Participation found the most common destination for travel for remote communities in the Northern Territory was Alice Springs, with the primary reasons for travel to be food/groceries and banking (Dockery 2015; A. Spandonide 2015a; Farrell 2006). Future work might focus on locationally diverse on-ground accounts of destination, frequency and purpose in order to better represent remote communities travel profiles and we caution here that without understanding critical details such as traveller motivation there remains a risk of designing an impractical network of hubs that doesn't, and will never, reflect real travel behaviours.

### *Electric Vehicles Charging Levels*

While thresholds defining electric vehicle charging levels vary across international jurisdictions in Australia the following definitions serve to describe charging systems, categorised across three levels, based on their power rating (Tran, Sutanto, and Muttati 2017; Herron 2016; Bräunl et al. 2020).

#### *Level 1*

Level 1 charging uses single phase power of up to 10A – often from commonplace General-Purpose Outlets to provide power up to 2.4kW (Bräunl et al. 2020; Queensland Government 2018b). The associated waiting time for a full charge can exceed 24 hours for a 60kWh battery.

#### *Level 2*

Level 2 charging is characterised by currents between 10-32A to supply power at 11-22kW (Queensland Government 2018b). These chargers typically require three-phase power supplies (Electric Vehicle Council 2020; Energeia 2018).

#### *Level 3*

Level 3 charging, also referred to as fast or ultra-fast charging, uses a Direct Current (DC) connection between vehicle and charger to deliver power at 25kW and above. These significant power ratings often require connection at a medium or high voltage at the level of the distribution network rather than the low voltages common at residential and business customers connection.

*Existing Charging Network*

EV enthusiasts have proven that it is already possible to drive around Australia in an EV using the current charging network (Tesla Owners Australia 2020; Tesla owners 2018). Approximately 17,000km of Australia's road network has been surveyed by the website PlugShare (Plugshare 2020) capturing available public charging sites including their physical access, electricity supply and plug type. The PlugShare website shows that some level of charging is available along most major highways of the National Land Transport Network running within and connecting Western Australia, the Northern Territory & Queensland (Infrastructure Australia 2019). While the current network may be suitable for non-time critical or recreational travelling there exist significant gaps between public charging points in Central Western Australia, Northern Territory & Western Australian border region, Northern Territory & Queensland border region (including Arnhem Land) and Central & Far North Queensland. Despite a single fast charger being installed in Darwin, and the Queensland Electric Super Highway providing 31 fast-charging sites on the densely populated East Coast between Cairns and Coolangatta (and west to Toowoomba), most charging points in remote Australia are either Level 1 or 2 necessitating charging for many hours or overnight (Plugshare 2020; Queensland Government 2018a; Infrastructure Australia 2019a).

**Methodology**

This section offers an overview of the data and methods that this paper draws upon to look at the road network travelable in relation to charging stops using two vehicle ranges, of 336km and 660km. Here we outline our methodology for defining service hubs, the collection and importation of data for the subsequent road network analysis and the rationale guiding initial selection thresholds used to select our sample high and low range electric vehicles.

### Defining Service Hubs

*Accessibility Remoteness Index of Australia (ARIA+)*

ARIA+ is used to measure the relative remoteness of Australian statistical areas known as Urban Centre Populations (UCPs). Remoteness of a locality is determined by first measuring the road distance to the nearest Service Centre in each of five categories (Categories A-E, shown in Table 1).

| Service Centre Category | Urban Centre Population |
|---|---|
| A | 250,000 persons or more |
| B | 48,000 - 249,999 persons |
| C | 18,000 - 47,999 persons |
| D | 5,000 - 17,999 persons |
| E | 1,000 - 4,999 persons |
| F (ARIA++ only) | 200 - 999 persons |

*Table 1: ARIA+ Service Centre Categories and Population Thresholds (Hugo Centre for Population and Housing 2016)*

Each distance value per locality is divided by the national average for that category to determine its standardised value. The standardised value is capped at 3 to avoid limit remoteness index values. The 5 standardised values are summed to gain the ARIA+ index out of 15 for each locality. Localities with scores (>5.92 – 10.53) & (>10.53 – 15) are considered "Remote" & "Very Remote" respectively. Figure 1 shows the ARIA+ 2016 results Australia wide.

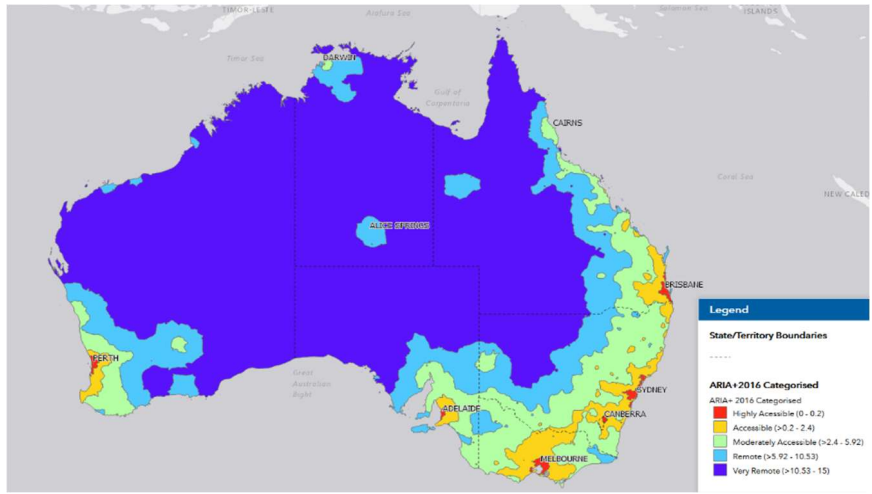

*Figure 1: ARIA+ 2016 Results Visualised  (University of Adelaide 2018)*

Category thresholds are justified as follows (Hugo Centre for Population and Housing 2016):

> *"Populated localities with populations of greater than 1,000 persons are considered to contain at least some basic level of services (for example health, education, or retail and as such these towns and localities are regarded as Service Centres"*

The Australian Statistical Geographic Standard (ASGS) uses ARIA+ results from each locality and overlays the ASGS Statistical Area Level 1 (SA1) data (Australian Bureau of Statistics 2016). This allows the Australian Bureau of Statistics (ABS) to define and publish remoteness statics for all States & Territories without gaps.

The assumptions based on these sources inform the experimental study that uses quantitative methods to collect data in order to model the basis for one possible solution to assessing the feasibility of EV travel by remote communities to larger service hub towns. This approach includes ranking remote communities into three population tiers:

large service hub towns, small service hub towns and non-service hub towns (including discrete Aboriginal communities) as well as defining their geolocation.

*Geographic Information Systems (GIS) - Road Network Analysis*

GIS are used across various disciplines to analyse and visualise spatial patterns to derive meaning for a given context (Rogers and Staub 2013). One useful application is to combine demographic statistics, remoteness and populations, represented as discrete points, and study their interactions with geographic spatial objects (such as road networks). In this study we used Quantum GIS (QGIS), which has a suite of network analysis functions for calculating the spatial connectedness of discrete remote localities and their populations on a given road network map (Rogers and Staub 2013; QGIS 2020; Raffler 2018). The implications for EV travel can be studied from the statistics derived from simulations using these functions.

A comprehensive set of remote towns with geolocation and population was compiled from the following three data sources:

1. Australian Government Indigenous Location (AGIL) dataset which provides the names and geolocations of all Indigenous communities across Australia.
2. Urban Centres & Localities (UCL) data structure from ABS census (Australian Bureau of Statistics 2016). This provides a list of urban centres (towns) and geolocations with population greater than 200. We refer to these UCLs as communities.
3. Population counts for these locations taken from the Census Tablebuilder (Australian Bureau of Statistics 2020). These provide Estimated Resident Populations (ERPs), which we refer to simply as populations.

These data sets are combined to create a single set of communities with geo-coordinates and population counts used in subsequent simulations. ARIA service centre population ranges, shown in Table 1: ARIA+ Service Centre Categories and Population Thresholds Table 1 were applied to define service hub and non-service hub localities. Specifically, communities with populations less than 1000 (Category F), are considered non-hub towns that make up origin destinations, communities with population 1000-4999 (Category E) are considered small service-hubs, while communities with population greater than 5000 are considered large service hubs.

Geoscience Australia TOPO250k road network provides a vector map on which network distances can be calculated (Geoscience Australia 2003) and these two datasets are imported into the QGIS environment where all network calculations and visualisations are performed.

*Network Analysis:*

The road travel distance between non-service hubs and service hubs is calculated using native and third-party network analysis tools in QGIS 3.12.2-Bucureşti. For the purposes of the experimental study an imported locality point file was filtered to remove localities south of <-28.00° proximal to a line running from Thargomindah in the east via Oodnadatta and Mimili on the Anangu Pitjantjatjara Yankunytjatjara (APY) Lands in the north-west of South Australia, via Leinster in the northern Goldfields to Kalbarri on the mid-west coast of Western Australia, including one state (Brisbane) and one Territory (Darwin) capital city (Geoscience Australia n.d.).

The hub population limits are applied to split the dataset into two subsets defining service hub locations and non-hub locations. An origin-destination matrix function, provided by third-party plug-in QNEAT3 (Raffler 2018), calculates the road distance for all origin-destination combination pairs. This function outputs a matrix with each combination and its associated road distance. The origin-destination matrix is queried using the Database Manager tool to determine the shortest distance and associated destination for each origin point. This queried matrix is joined to the original point layer to create a new layer of origin points with destination and distance values appended. Each non-hub location now has a defined network distance to its nearest hub.

Using the Data Management Toolbox, the new origin point layer is split by hub accessibility (i.e. distance to hub is zero, 1, 2, 3 EV ranges), to group localities by the number of recharging stops required. Once grouped they are exported to calculate the relevant statistics regarding population and the number of towns within range of a given number of recharging stops, for further analysis. The road network travelable with 0 to 3 charging stops from hubs is visualised using the Service Area (from layer) algorithm in the network analysis toolbox. The 'Join by Lines' algorithm is used to create and visualise hub lines connecting hub and nearest non-hub localities. Finally, the map with layered data is then exported for four separate simulations whereby small and large hub size utilizing high and low vehicle ranges are applied.

*Vehicle selection and range calculations*

Our study considered those five-seater EVs available in Australia in August-2020 and selected from the highest and the lowest range models having a published range of greater than 300km (EV Council 2020). This selection threshold is based on a preliminary calculation of average travel distances from discrete "non-service hub" to "service hub" communities in remote Australia, as shown in Figure 2.

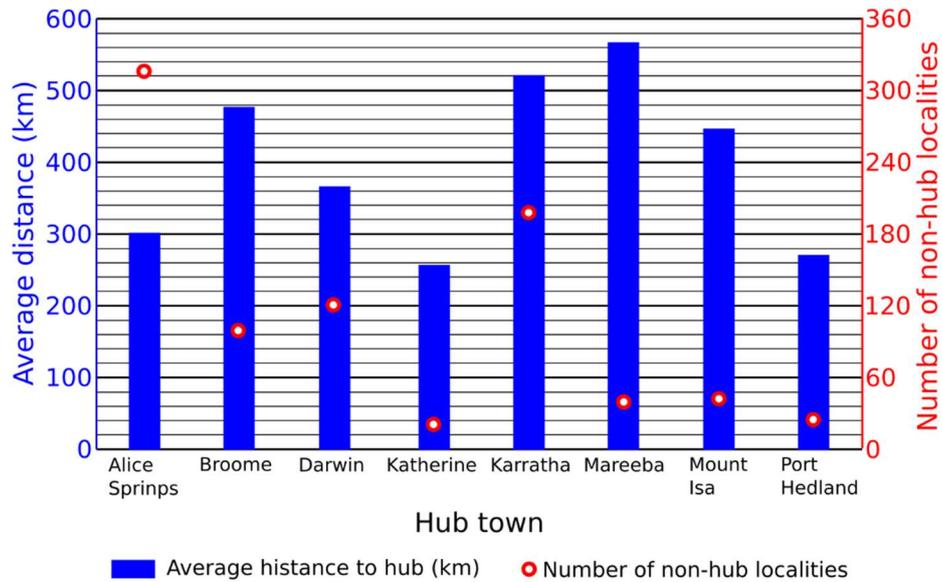

*Figure 2: Average Travel Distances from non-service hub to closest service hub town.*

In determining an initial selection threshold, we note that only three localities - Alice Springs, Port Hedland and Karratha - are reachable by a "non-service hub" town with a range of less than 300km on average.

| | Specifications | Tesla Model S | Audi e-tron 50 Quattro |
|---|---|---|---|
| **Electric vehicle specifications** | Battery Size (kWh) | 100 (Electric Vehicle Council 2020) (Jet Charge 2020) | 71 (Audi Australia Pty Ltd 2020) |
| | Onboard Charger (kW) | 11 (Jet Charge 2020) | 11 (Audi Australia Pty Ltd 2020) |
| | Electric Range (km) | 660 (Electric Vehicle Council 2020) (Jet Charge 2020) | 336 (Electric Vehicle Council 2020) |
| | **Charging Type** | | |
| **Estimate recharge times** | Level 1: 2.4kW | 43h 28m (Jet Charge 2020) | 30h 52m |
| | Level 2: 11kW | 9h 5m (Jet Charge 2020) | 6h 27m |
| | Level 3: 50kW | 2h (Jet Charge 2020) | 1h 25m |

*Table 2: Selected specifications for sample vehicles Audi e-tron 50 Quattro & Tesla Model S and approximated charging times.*

Noting that our testing scenario applies one vehicle type for all journeys per simulation we selected from among those vehicles with a published range of greater than 300km - a longer-range of 660km (Tesla model S) and a lesser range of 336km (Audi e-tron 50 Quattro) - for the following simulations (Tesla Inc 2020; Jet Charge 2021; EV Council 2020)[1]. The relevant vehicle specifications are shown in Table 2 as are estimated charging times using three of the more commonly available charging infrastructures. Batteries are assumed to be charging from 0% – 100% while the wait times associated with a given charging type are calculated by dividing the battery capacity by the load power of the charger[2].

---

[1] Vehicle ranges accurate at time of citation (August 2020). Vehicle ranges expressed in All Electric Range WLTP.

**Analysis**

Here we analysed two scenarios defined by the size of the service hub travelled to: large service hubs with estimated resident populations of more than 5000 residents, and small service hubs with populations of more than 1000 residents (inclusive of large service hubs). Within each hub scenario, we considered two electric vehicle classes: those with a lower range of 336km, and those with a longer range of 660km. For each vehicle and hub scenario the number of communities and residents that could reach their nearest service hub using an electric vehicle stopping either zero (direct), one, two or three times to recharge were calculated, shown visually using maps labelling the number of recharging stops for each community.

*Scenario 1: Large Service Hubs (population > 5000)*

Scenario 1 considers towns with population greater than 5000 as "service hub" towns. communities of less populations than 5000 are in this scenario defined as "non-service hub" towns and considered origin points for trips to service hub towns.

*Scenario 1.1: Lower-range vehicle (336km)*

Figure 3 shows the colour coded non-hub towns requiring zero (direct), 1, 2, and 3 recharge stops to access their nearest large service hub using a 336km range vehicle.

---

[2] Onboard Charger capacity in Table 2 caps the amount of DC power receivable by the EV when attempting to charge with AC power from the grid or any other standalone system [56].

The Hub-lines are shown connecting non-hub towns with their closest associated hub town while service areas denoting the maximum road distance travelable with zero (direct), 1, 2 & 3 charging stops surround each hub, visualised as coloured portions of the corresponding road network.

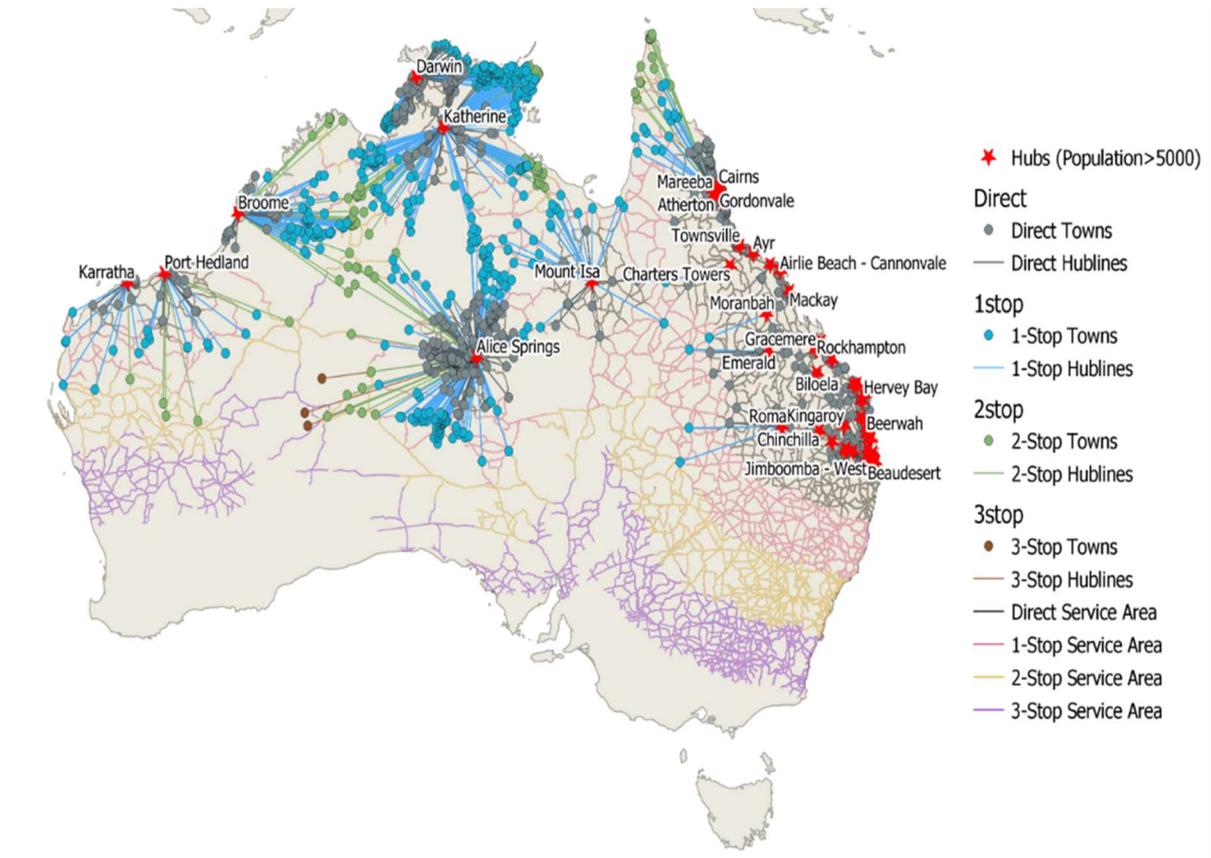

*Figure 3: Network map of non-service communities and their closest large service hub (population>5000) using a lower-range EV (336km range).*

A statistical summary of the level of accessibility of populations to their corresponding defined service hub towns using the lower-range EV is provided in Table 3, showing that 370,340 (80.73%) non-hub residents residing in 636 localities would have direct access to a service hub. This suggests that most remote communities live within direct road access of those primary services they require. This is important as localities with direct access may feasibly only require charging at their origin and destination locations, meaning the barriers associated with establishing roadside charging infrastructure for these communities are potentially avoided while the utility of establishing or building upon infrastructure that facilitates home-charging benefits communities directly. An additional 66,811 residents (14.56%) in a further 448 localities would require a single charging stop to reach their closest service hub. In the presence of suitable recharging locations, a single recharging stop may or may not be considered prohibitive (or impractical) by remote living residents. The remaining 101 localities are so distant from large service hubs that their populations (21,431) would require two or more recharging stops to reach their nearest service hub. In the absence of Level 2 or Level 3 fast charging infrastructure this would add several days of charging to each trip likely making travel prohibitive using lower-range EVs.

| Hub Population | Vehicle Range | Direct Access | | 1-Stop | | 2-Stop | | 3-Stop | |
|---|---|---|---|---|---|---|---|---|---|
| | | Towns (% of non-hub communities) | Population (% of non-hub population) | Towns (%) | Population (%) | Towns (%) | Population (%) | Towns (%) | Population (%) |
| **>5000** | 336km | 636 (52.71%) | 370,340 (80.73%) | 448 (37.84%) | 66,811 (14.56%) | 88 (7.43%) | 21,366 (4.66%) | 3 (0.25%) | 65 (0.014%) |

*Table 3: Scenario 1.1 results: Accessibility of low-range (336km) vehicle to large service hubs by number of communities and associated potulations.*

*Scenario 1.2: Longer-range vehicle (660km)*

Figure 4 shows the colour coded non-hub towns requiring zero or a single charging stop when utilizing a longer-range (660km) EV in the large hub size scenario, illustrating an extended service area requiring zero or one stop and a significant increase in the number of communities directly accessible to their service hub in this scenario. The statistical summary in Table 4 shows that almost all (>90%) remote communities and their populations in this scenario have direct access to their service hub town assuming the use of the longer-range EV.

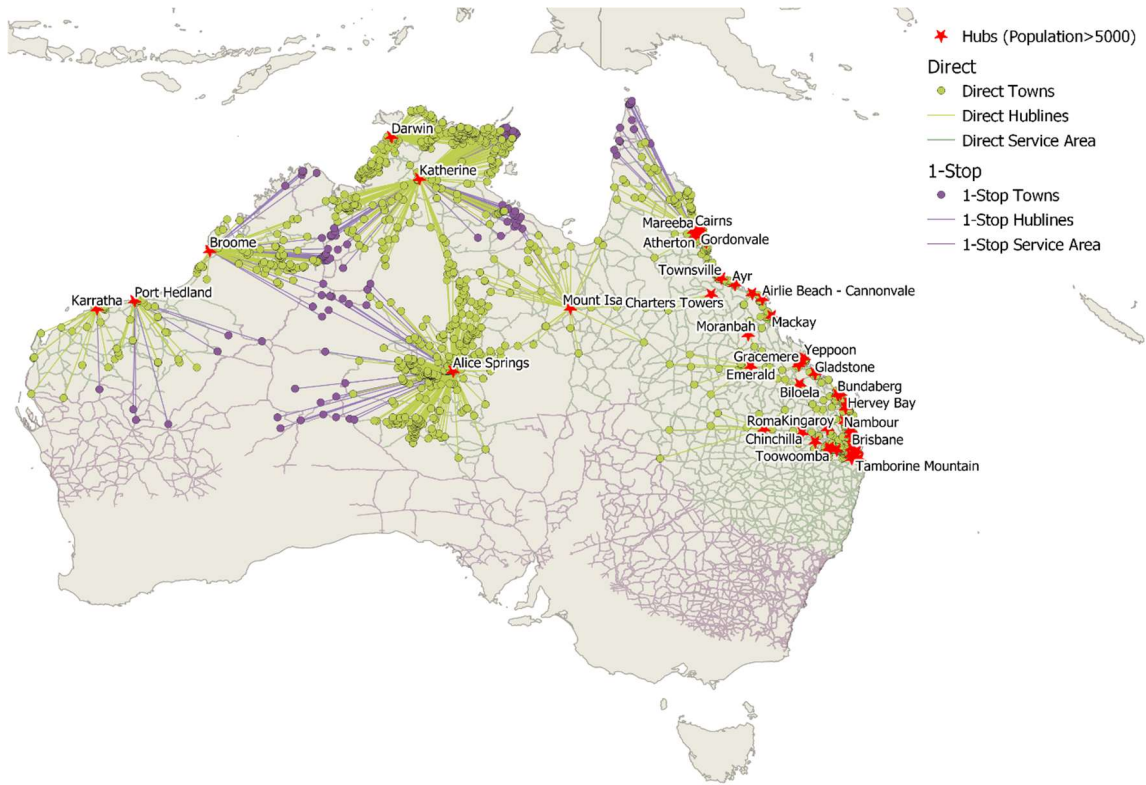

*Figure 4: Network map of non-service communities and their closest large service hub (population>5000) using a longer-range EV (660km range).*

The direct service area shown in green in Figure 4 is reachable by all discrete non-hub communities without stopping, requiring charging only at origin and destination locations. In this scenario 8.9% of non-hub towns (with a population totalling 19,533) would have to factor recharging en-route as part of their travel plans. Consequently wait-times would be longer than for the low-range vehicle scenario due to the larger battery capacities of the long-range EV. While the context of travel is crucial in assessing what a reasonable hub charging wait-time is and whether partial recharging en-route is considered practicable, Level 3 charging for en-route stopping would be recommended for safety and convenience.

Accessibility to services in this scenario requires less investment in the charging network both at hubs and en-route locations. Relatively few people are outside of direct hub access and those who are could potentially partially charge en-route in order to reach their relevant hub. As no towns lie outside of one stop range, the greater inconvenience experienced with multiple stops does not apply. This scenario presents a reasonable level of feasibility for remote access to hubs – provided en-route infrastructure exists and is of an appropriate charging speed to match the motivations of travellers. Level 3 charging would require only 1.5hrs waiting, however we note there are very limited en-route fast chargers available outside of the major centre of Darwin (Plugshare 2020). Level 2 charging is the best cases scenario currently available en-route and requires a wait-time of at least 6 hours. Level 1 is not feasible unless the circumstances are unique in that a multi-day stopover is both acceptable and can be safely achieved.

| Hub Population | Vehicle Range | Direct Access | | 1-Stop | |
|---|---|---|---|---|---|
| | | Towns (% of non-hub communities) | Population (% of non-hub populations) | Towns (%) | Population (%) |
| **>5000** | 660km | 1070 (90.04%) | 436,618 (95.18%) | 106 (8.9%) | 19,533 (4.26%) |

*Table 4: Simulation 1.2 results: Accessibility of longer-range vehicles to large service hubs by number of communities and associated populations.*

### *Scenario 2 – Small Service Hubs (population > 1000)*

The second scenario follows the same method as the first, with the exception that all communities with population greater than 1000 are designated service hub towns. Broadening the population threshold for hub classification sees 39 communities' origins converted to service hubs. By redefining the population cut- off the emergence of new important hub locations can be observed. As shown in Figure 5 new hub clusters appear in eastern and central Queensland, north-west Northern Territory (Arnhem Land), and north-west Western Australia.

These smaller hubs reduce demand on the larger hub towns of Alice Springs, Darwin & Katherine, each of which were previously the closest hub for a preponderance of remote communities. Yulara, Kununurra, Halls Creek, Tennant Creek, Maningrida, Nhulunbuy, Ngukurr are all notable examples.

### *Scenario 2.1: Lower-range Vehicle (336km)*

Assuming the use of low-range EVs, towns that previously required charging stops in Scenario 1 are now directly accessible to smaller hubs. A key example is shown in Figure 5 whereby some towns, between and proximate to the area defined by nodes Alice Springs, Mt Isa, Katherine and Broome, are now redirected to a suite of smaller hubs, primarily Doomadgee, Tennant Creek, Ngukurr, Kununurra, Halls Creek & Fitzroy Crossing.

By reducing the service hub population cut-off to 1000 population, direct hub access has now increased from 52.71% to 89.64 % of non-hub towns. This equates to an additional 77,720 residents gaining direct access (zero stops) to hubs under this scenario. As direct hub access implies that charging only takes place at the origin and destination locations, the need for en-route charging may diminish while the necessity for various levels of improved charging infrastructure at both home and hub locations remains critical. Depending on the specific context of the required travel, including the time spent at destination and the frequency of travel, all levels of chargers may be considered adequate (if not ideal) for both home and destination. The longer wait-times typically associated with Level 1 charging become less prohibitive as longer duration stays are typically more common at home and destination than en-route. Assuming then that vehicle demand is lesser once the traveller arrives at either location, the acceptability of waiting days until a vehicle is fully charged will depend upon the interval until the next intended travel.

Of remaining residents, the majority (10,057) would require one stop when traveling from their residence in one of 201 communities, to their nearest service hub. Only 6 communities, with a total estimated resident population of 465, would require two charging stops under this scenario. As before, performing one charging stop – particularly of the lower-range EV – may be feasible, while two or more stops may be a prohibitive factor.

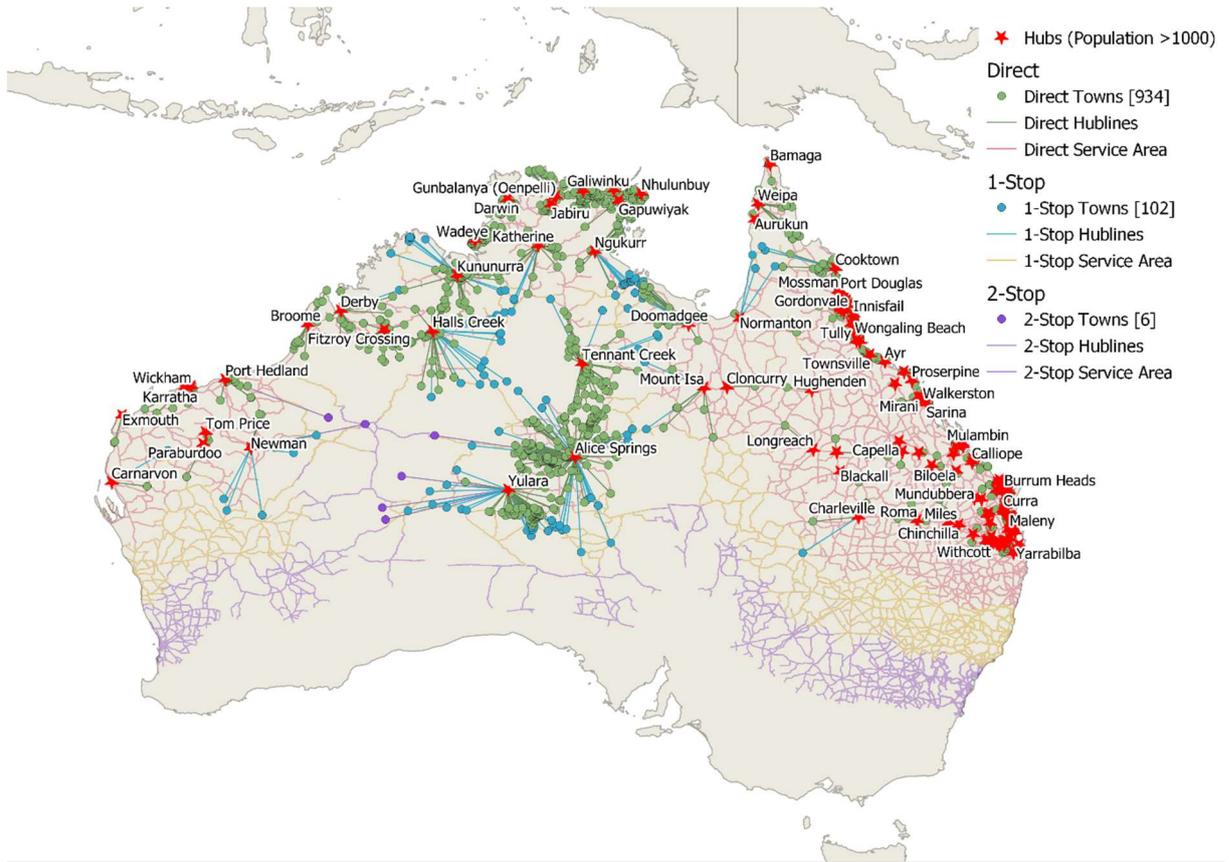

*Figure 5: Network map of non-service communities and their closest small service hub (population>1000) using a lower-range EV (336km range).*

| Hub Population | Vehicle Range | Direct Access | | 1-Stop | | 2-Stop | |
|---|---|---|---|---|---|---|---|
| | | Towns (% of non-hub communities) | Population (% of non-hub populations) | Towns (%) | Population (%) | Towns (%) | Population (%) |
| **1000** | 336km | 934 (89.64%) | 370,340 (93.25%) | 102 (9.78%) | 10,057 (6.45%) | 6 (0.58%) | 465 (0.30%) |

*Table 5: Scenario 2.1 results: Accessibility of low-range vehicles to small service hubs by number of communities and associated populations.*

*Scenario 2.2: Longer-range vehicle (660km)*

The use of longer-range vehicles to access service hubs of greater than 1000 population provides the greatest level of access, with 99.42% of non-hub localities – representing 99.70% of the population – having direct access. This is shown in Figure 6 and Table 6.

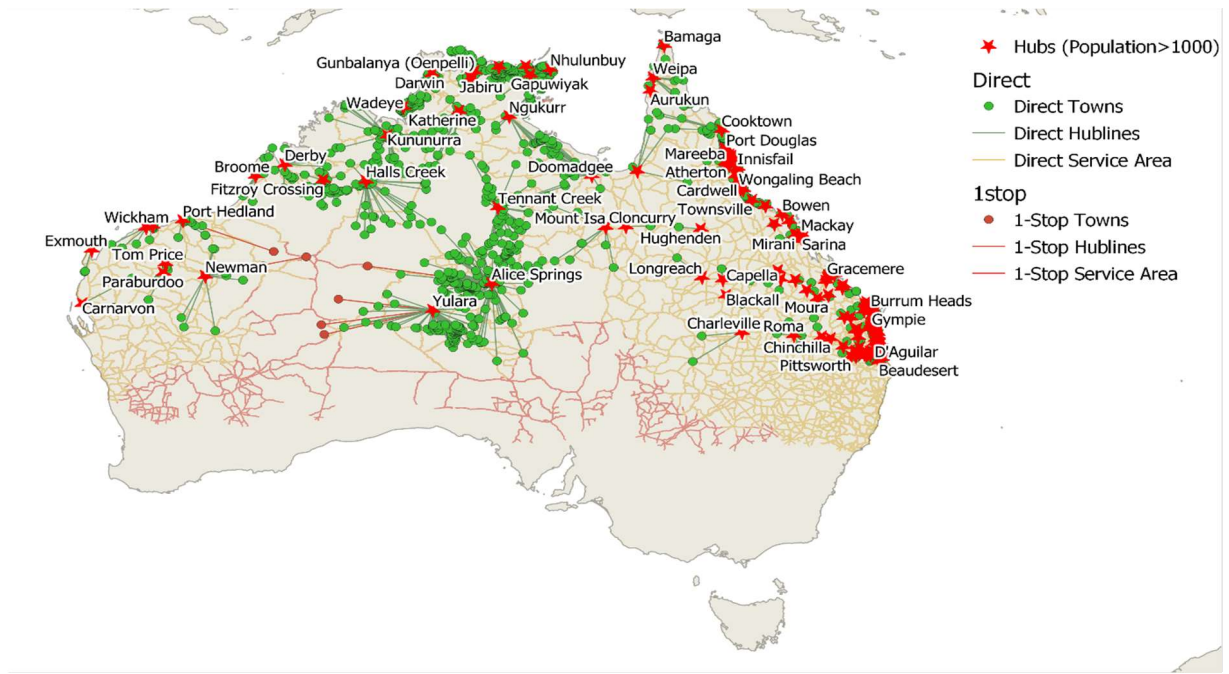

*Figure 6:  Network map of non-service communities and their closest small service hub (population>1000) using a longer-range EV (660km range).*

Applying vehicles with a longer range unsurprisingly results in fewer charging events with almost zero en-route charging points imperative to access hub towns directly. Charging may only be required at origin and destination locations and full (100%) recharges are no longer strictly required as non-hub towns are generally lesser than the full stated vehicle range in total distance from their hub destination. This has the potential to reduce both load on hub chargers and the necessity for fast charging, as replenishing the total capacity of the vehicle battery is no longer a critical requirement.

| Hub Population | Vehicle Range | Direct Access | | 1-Stop | |
|---|---|---|---|---|---|
| | | Towns (% of non-hub communities) | Population (% of non-hub populations) | Towns (%) | Population (%) |
| **<1000** | 660km | 1036 (99.42%) | 155,549 (99.70%) | 6 (0.58%) | 465 (0.30%) |

*Table 6: Scenario 2.2 results: Accessibility of longer-range vehicle to small service hubs by number of communities and associated populations.*

With no en-route charging events this scenario presents an idealised vehicle range and charging network scenario. Should only Level 1 charging be present at hub and non-hub locations, the waiting time until fully charged would be approximately 43 hours. Depending upon travellers demands and their willingness or desire to spend days at hub locations, this may or may-not be considered an acceptable wait time however if faster charging technology is available, trips to service hubs become more practically feasible. Level 2 charging requires a maximum wait time of up to 9 hours - however should travellers wish to make same-day return trips access to Level 3 charging becomes imperative. Similarly, for the <1% of towns (and their resident populations) requiring en-route charging, Level 3 services are most desirable, while Level 2 en-route charging may or may not be feasible depending upon the specific circumstances of the traveller and their stopped location.

Having fewer en-route charging events suggests that necessary charging is largely performed at home or hub location. Given the larger battery capacities of longer-range EVs this has the consequence of increasing energy demand at home and hub locations, and we caution that supply-side investment in support of increased energy demand from charging comes with its own set of planning and access challenges.

**Implications**

Our results demonstrate that the utility of electric vehicles in remote communities is perhaps more feasible than might at first be expected. Even within the limited range of those EVs currently available in Australia, a majority of intended trips between communities and service hubs could be travelled without recharging en-route. This suggests that the barriers to EVs is not solely the oft-quoted tyranny of distance between remote towns, or the range of existing EVs, but a combination of the availability of charging infrastructure in towns, communities and en-route charging stations plus likely a multiplicity of more prosaic barriers including accessibility, availability and financing of EVs in remote markets, the training of technicians (and lay people) to service and repair EVs, and perceptions related to the use case of EVs in remote areas.

The major implication of this finding is that remote communities ought to be incorporated early into planning for the electrification of transport– through initiatives such as the "Preparing the Northern Territory for Electric Vehicles" process (Hunt et al. 2019). Our results for small service hubs (population > 1000) suggest that while it may be expedient to focus on installing Level 2 and 3 charging infrastructure in settlements and hubs, this may be complicated by factors including the through-put of non-resident traffic such as freight at (unpopulated) en-route locations.

Deploying and operating such charging infrastructure is a complex challenge. Barriers likely include weak power supplies in many communities and the presently non-existent (or standalone) power supply at intermediate locations such as roadhouses. The required power rating of chargers will be influenced by the finer details of travel patterns and the charging times that these make feasible, but irrespective, powering chargers within towns and at intermediate locations will likely pose a challenge as many of these locations are supplied electricity from standalone diesel or gas fired power systems sized to cover existing (historical) loads ranging from hundreds of kilowatts to tens of mega-watts (AECOM 2014). In this context, the addition of a single Level 2 charger may add the equivalent of multiple households of unplanned and unprecedented extra loading. "Thin" distribution networks may require upgrading to handle additional demand, as well as the potential voltage fluctuations and harmonic distortion that EV chargers introduce (Clement-Nyns, Haesen, and Driesen 2011). To address these issues and facilitate additional charging loads, grid infrastructure upgrades would likely be required in many circumstances (Tran, Sutanto, and Muttati 2017). An alternative approach is the installation of standalone power systems and microgrids powered by solar and batteries. These are well suited to remote contexts, where their generation can either reduce the load on distribution networks or fully replace fossil fuel generation systems, and are gaining traction with deployments in remote Australia (Energy Networks Association 2022; Horizon Power 2022; Department of Industry 2021). Because standalone power systems and microgrids are often designed custom for a given site and costly to modify, it is important to that their design process begins to incorporate serving EV charging loads in the medium-term future, either by oversizing their current design or having clear options for future expansions.

**Conclusion**

This experimental study aimed to assess the feasibility of electric vehicle travel for remote communities using hub and spoke accessibility scenarios. Our results indicate that the uptake of electric vehicles for remote communities is technically a possibility in an idealized scenario, as the vast majority (93%) of residents would have direct access – without needing to recharge en-route – to their nearest small service hub (population>1000) with even the lower-range of currently available EVs in the Australian market (with a range of 336km). Furthermore, 95% of residents would have direct access to their nearest large service hub (with population>5000) with a currently available, longer-range EV.

These results indicate that the hub-and-spoke model may serve as an appropriate model for deploying EV charging infrastructure with a focus on installing Level 2 and 3 chargers in small and large service hub towns, rather than only at en-route roadside locations. This is particularly attractive given the current absence of power supplies along roads and the significant wait times to fully charge EVs, together with extant challenging weather conditions that increase the importance of co-locating hospitality services for travellers.

Our study uses simplifications that require further research to refine. Chief among these is the need to quantify the impact of unsealed road conditions and temperature extremes on the effective range of EVs. Future research that quantifies these effects would have major impacts for the electrification efforts in remote Australia and across the Global South. In the acute circumstance where these impacts have the effect of halving stated vehicle range - our results for lower-range EV would be achieved by the long-range EV. The second major idealisation is that the power needs of EV charging could possibly be serviced by existing power systems in communities. In origin communities it may be feasible to use solely Level 1 charging – minimising the stress placed on community power supplies – but in small and large service hubs Level 2 or 3 charging will be required to enable return or onward travel on the same day. These higher power levels of charging may be a major stress on the extent weak power systems in remote Australia, including those of large service hubs. Other issues to consider include the operating models for public charging stations and what valuable services may be lost with the closure of existing petrol stations.

Lastly, we stress that the desktop, statistical approach taken in this paper glosses over significant variability in local contexts and travel behaviours in remote Australia. Future research could use questionnaires and interviews to form a more holistic and multi-dimensional understanding of current and evolving travel patterns and the services desired at EV charging places. Matching these needs and preferences will be essential for effective planning of infrastructure and services. Like all new technology, EVs and EV charging must be deployed with respect for local circumstances and with the meaningful and engaged participation of remote living residents in order to prioritize the adoption of new technologies to residents' wants and needs.